\documentclass[twocolumn, showpacs, amsmath,amsfonts,prb]{revtex4}
\usepackage{graphicx}

\begin{document}

\title{Coupled electron--heat transport in nonuniform thin film semiconductor structures}
\author{V. G. Karpov} \affiliation{Department of Physics and Astronomy, University of Toledo, Toledo, OH 43606, USA}

\begin{abstract}

A theory of transverse electron transport coupled with heat transfer in semiconductor thin films is developed conceptually modeling structures of modern electronics. The transverse currents generate Joule heat with positive feedback through thermally activated conductivity. This can lead to instability known as thermal runaway, or hot spot, or reversible thermal breakdown. A theory here is based on the  optimum fluctuation method modified to describe saddle stationary points determining the rate of such instabilities and conditions under which they evolve. Depending on the material and system parameters, the instabilities appear in a manner of phase transitions, similar to either nucleation or spinodal decomposition.

\end{abstract}
\pacs{72.60.+g, 72.80.Ng, 64.60.Q-, 73.50.Fq}
\maketitle
\date{\today}

\section{Introduction}{\label{sec:intro}
Various treatments of electronic transport in disordered systems typically concentrate on systems at a given fixed temperature.
However, observations (see references below) often  point at the coupled electron-heat transport where local  fluctuations in electric current generate temperature fluctuations. When the latter have positive feedback, as e. g. in the case of thermally activated conductivity, an instability arises leading to the current filamentation. `Weak spots' corresponding to suitable disorder configurations promote such instabilities. While this mechanism has long been known qualitatively, \cite{subashiev1987} its quantitative understanding remains insufficient leaving open questions about the role of material and structure parameters, and effects of static vs. thermodynamic fluctuations.

This work attempts a theory of coupled electron-heat transport concentrating on a rather representative case of transverse conduction through thin-film structures.  A model structure consists of an active (heat generating) conducting layer between two electrically  inactive insulating layers representing encapsulation always found with electronic devices. This structure is depicted in Fig. \ref{Fig:sample}. The active layer can be a single or multi-layered semiconductor sandwiched between thin metal electrodes. The electric potential along each of the electrodes is constant; the potential difference $V$ between them is maintained by an external power source.

The disorder is introduced through the activated transversal electric conduction with random Gaussian activation barriers varying in the lateral (along the film) directions. The role of insulating layers is that they affect the temperature distribution and make the entire model more realistic. For simplicity, we assume one of them totally insulating while another one having a finite thermal conductivity. Also, for simplicity, thermal conductivities and specific heats of the active and insulating layers are assumed the same.

\begin{figure}[tb]
\includegraphics[width=0.37\textwidth]{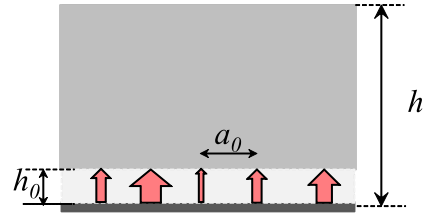}
\caption{Sketch of the system with nonuniform power generation and current flow (fat arrows). Top insulating layer shown in gray. The bottom layer (dark) represents a strong thermal insulator.  \label{Fig:sample}}
\end{figure}

The analysis below is aimed at finding the probability of local temperature fluctuations and their radii associated with locally increased current density vs. the system dimensions, material parameters, and ambient temperature. It is based on the premise of localized rare lateral fluctuations that do not overlap. These localized entities are similar to other types of localized  states in disordered systems, for which theoretical description known as the optimum fluctuation method (OFM) has been developed long ago. OFM was originally created to describe electronic states in band tails of disordered semiconductors; \cite{lifshitz1964,halperin1968,zittarz1967,lifshitz1987,meighem1992} it was applied later to localized sound excitations in glasses, \cite{karpov1993a}, resonance electronic states in disordered metals, \cite{karpov1993b, apalkov2002a}  fluctuation tail states in magnetic semiconductors, \cite{karpov1994} random lasing in disordered dielectric films, \cite{apalkov2002b}, local fluctuations in thermal expansion of glasses, \cite{karpov1992} and nucleation in disordered media. \cite{karpov1996}

The essence of OFM is in the optimization of configurational probability (or entropy) of fluctuations under the additional condition that the dynamical characteristic of a fluctuation satisfies the appropriate differential equation (Schr\"{o}dinger equation for electronic state, elastic wave equation for sound excitations, electromagnetic wave equation for optic modes, etc.). This is achieved through the variational approach, in which the dynamical characteristic is kept fixed (yet arbitrary) in the course of optimization of the configurational entropy, after which it is optimized to additionally minimize that entropy. The details of OFM vary between different systems. Here developed OFM is tailored to describe the temperature fluctuations coupled with the electric current, so that the dynamical characteristic (temperature) of fluctuations satisfies the heat transfer equation.

The analysis below shows that hot spot instabilities evolve in a manner of phase transformations, either by nucleation or similar to spinodal decomposition affecting the entire area. The nucleation scenario of such instabilities in uniform systems was established earlier based on general phenomenological analysis. \cite{subashiev1987}

This paper is limited to a general theoretical analysis; possible applications of the coupled electron-heat transport will be presented in more appropriate journals. We refer to a recent monograph \onlinecite{vashchenko2008} for many practically important cases. The relevant observations are found with bipolar transistors \cite{vashchenko2008,liou1996,olsson2001,breglio2000,bagnoli2009}, other metal-insulator-semiconductor structures, \cite{semenov2006,tsan2002,bolte1998,alagi2011,choi2011,wang2009}, nanoscale transistors, \cite{pop2006}, graphene transistors, \cite{bae2011}, and thin-film photovoltaics.\cite{radue2010,meyer2004,mcmahon2005,shvydka2004} In these applications, the phenomenon under consideration was labeled as thermal runaway, or hot spot, or (reversible) thermal breakdown. It can be detrimental to the corresponding device operations leading to their irreversible degradation in hot spots via local shunting, burning, or melting; hence, significance for device reliability.

The paper is organized as follows. Sec. \ref{sec:transport} introduces the basic equations describing the coupled electron-heat transport in a non-uniform system. To better explain the essence of OFM and subsequent results, two toy models are considered in Sec. \ref{sec:toymodel}. Sec. \ref{sec:OFM}, presents a modification of OFM  describing saddle points through which the system evolves into thermally non-uniform state. The OFM functional is optimized in Sec. \ref{sec:variational} through direct variational procedure. The steady state rate of hot spot nucleation is estimated in Sec. \ref{sec:FPE}. Finally, Sec. \ref{sec:disc} presents general discussion and conclusions.

\section{Coupled electron and heat transport in a disordered system}\label{sec:transport}

The Joule power density is given by
\begin{equation}
P=P_0\exp(-E/kT), \quad P_0={\cal E}^2\sigma _0\exp\left(-\frac{\overline{E}}{kT}\right).
\label{eq:locpow}\end{equation}
Here ${\cal E}=V/h_0$ is the electric field strength where $h_0$ is the distance between the electrodes (see Fig. \ref{Fig:sample}). $\sigma _0$ is the pre-exponential of conductivity,
$$\sigma = \sigma _0\exp[-(\overline{E}+E)/kT]$$
with $\overline{E}$ being the average activation energy, $k$ is the Boltzmann's constant, and $T$ is the local temperature.
The random part of activation energy, $E$ has zero average, $\langle E\rangle =0$ and a finite dispersion $\langle E^2\rangle =B$. It is characterized by the correlation function
\begin{equation}
\langle E({\bf r})E({\bf r'})\rangle =Bs\delta ({\bf r-r'}).
\label{eq:corfun}\end{equation}
Here the radius vector ${\bf r}$ lies in the film plane, $\delta ({\bf r})$ is the two-dimensional delta function implying zero correlation radius disorder. The minimum area $s$ is determined by the physical nature of fluctuations. For example, its characteristic linear scale $a_0\sim s^{1/2}$ (likely in sub-micron range) can be given by the screening radius or the grain size, or other length, below which the system parameters do not vary significantly. $s$ is introduced to give $B$ the dimensionality of the square of energy and the meaning of the dispersion of random energies $E$.

Local elements of the system interact through heat transfer described by the standard equation
\begin{equation}
\chi \nabla ^2T + P({\bf r}) = 0
\label{eq:heattr}\end{equation}
where $\chi$ is the thermal conductivity and the Laplacian $\nabla ^2$ is three dimensional, and $\chi$ is coordinate independent. The power generation density is a sum of average and random contributions,
\begin{equation}
P= \langle P\rangle +P^{(1)},\quad \langle P\rangle \equiv P_0\left\langle\exp\left(-\frac{E}{kT}\right)\right\rangle.\nonumber
\end{equation}
where
\begin{equation}
P^{(1)}= P_0\exp\left(-\frac{E}{kT}\right) -\langle P\rangle.
\label{eq:fluctP}\end{equation}

Eq. (\ref{eq:heattr}) assumes the steady state heat transfer. The assumption of stationary states is common to all known cases of OFM. The problem under consideration, however, is different with respect to the notion of stationary fluctuations. Since the instability  evolves in a fashion of phase transitions, the stationary solutions of Eq. (\ref{eq:heattr}) can only describe saddle points in the parameter space. The temperature fluctuation $\delta T$ becomes time dependent in the proximity of each of such point, described by
\begin{align}\label{eq:transheattr}
-C\delta T/\tau = \chi \nabla ^2T + P({\bf r})
\end{align}
in the relaxation time approximation, where $C$ is the specific heat. The fluctuation decay will correspond to positive, while fluctuation growth (instability) to negative values of $\tau$; this criterion is used in Sec. \ref{sec:variational} below.
\section{Linear approximation: no-breakdown steady state regime}\label{sec:linear}
For completeness, consider briefly a trivial situation where the disorder $B$ and temperature fluctuations $\delta T$ are small allowing the linearization
\begin{equation}
P=P_0\left[1+\frac{E({\bf r})}{kT_0}-\frac{\overline{E}}{kT_0^2}\delta T({\bf r})\right]\label{eq:linearappr}\end{equation}
where $T_0$ is the average temperature. Substituting this into Eq. (\ref{eq:heattr}) and setting $$\delta T(r,z)=\phi (r)\exp (z/z_0),\quad z_0=const$$ for the radial ($r$) and transversal ($z$) coordinates yields
\begin{equation}
\nabla _r^2\phi -\frac{1}{r_0^2}\phi=u({\bf r}).\label{eq:lineareq}\end{equation}
Here $\nabla _r^2$ is the two-dimensional Laplacian, $$\frac{1}{r_0^2}\equiv\frac{P_0\overline{\overline{E}}}{\chi kT_0^2},\quad \overline{\overline{E}}\equiv \overline{E}-\frac{\chi kT_0^2}{z_0^2P_0}, \quad u({\bf r})\equiv\frac{P_0E({\bf r})}{\chi kT_0}$$
and $z_0$ must be determined from the boundary conditions.
The solution to Eq. (\ref{eq:lineareq}) has the form
\begin{equation}\label{eq:hankel}
\phi ({\bf r})=(-1/4)\int d^2ru({\bf r^{'}})H_0^{(1)}(i|{\bf r-r^{'}}|/r_0)
\end{equation}
where $H_0^{(1)}$ is the Hankel function.

The quantity in Eq. (\ref{eq:hankel}) represents a sum of large number of random contributions and, according to the central limit theorem, is a random quantity itself with the Gaussian probability distribution. Its dispersion $\langle\phi ^2\rangle$ is given by
\begin{eqnarray}\label{eq:weakdisp}
&&\frac{1}{16}\int _0^{\infty}d^2r^{'}d^2r^{''}H_0^{(1)}\left(\frac{ir^{'}}{r_0}\right)H_0^{(1)}\left(\frac{ir^{''}}{r_0}\right)\langle u({\bf r^{'}})u({\bf r^{''}})\rangle \nonumber \\
&&=\frac{\pi r_0^2P_0^2Bs}{4\chi ^2(kT_0)^2}=\frac{\pi P_0Bs}{4\chi\overline{\overline{E}}k}.
\end{eqnarray}
Here we have taken into account Eq. (\ref{eq:corfun}) and the value \cite{watson} of the integral $\int _0^{\infty}[H_0^{1}(x)]^2xdx=2$.

To avoid unnecessary discussions of boundary conditions $z_0$ and $\overline{\overline{E}}$ are left as two parameters. Neglecting the temperature change through the active film ($\exp(h_0/z)\approx 1$), the net result is that the temperature fluctuations are characterized by the radii of $r_0$ and the Gaussian distribution,
\begin{equation}\label{eq:linearres}
\rho (\delta T)\propto \exp\left(-\frac{\delta T^2}{\delta T_0^2}\right)\quad {\rm with}\quad \delta T_0^2=\frac{\pi P_0Bs}{4\chi\overline{\overline{E}}k}.
\end{equation}

The important point is that the above linear approximation does not account for positive feedback of temperature fluctuations on transversal conduction and thus the disorder remains fixed and temperature independent. While this restriction eliminates the possibility of thermal breakdown (which is the main topic here), the results of this section can still be applicable to the case of very small currents and fluctuations used e. g. in thermography diagnostics. \cite{shvydka2004,breitenstein}

\section{Toy models}\label{sec:toymodel}
Because the regular OFM below is mathematically cumbersome, it is illustrated here with simplified  (toy) models. One of them concentrates on the case when there is no positive feedback on conductivity by local heating. Another one deals with a homogeneous system  and concentrates on the positive feedback.
\subsection{Conductive filaments through an insulating film}\label{sec:condfil}
Consider a two phase structure where transversal current flows through conductive filaments in an insulating host of thickness $h_0$ sandwiched between two equipotential electrodes. The structure is characterized by the average transversal conductivity $\overline{\sigma}$ due to filaments of average concentration $\overline{n}$ per area. Local fluctuations $\delta n$ in their concentration  result in the corresponding conductivity fluctuations  $\delta \sigma =\overline{\sigma}\delta n/n$. Since the filaments generate Joule heat, they create fluctuations $\delta T$ in temperature; the tail of probabilistic distribution of $\delta T$ is found below.

Consider a cylinder shaped region of radius $a$ perpendicular to the electrodes where the characteristic fluctuation in filament concentration is $\delta n$. The Gaussian probability of such a fluctuation is estimated as
\begin{equation}\label{eq:Stoy}\exp\left[-\frac{(\delta n)^2a^2}{\overline{n}}\right]=
\exp\left[-\overline{n}a^2\left(\frac{\delta \sigma}{\overline{\sigma}}\right)^2\right]
\equiv \exp(-S)
\end{equation}
$S$ can be optimized with respect to $a$ after $\delta\sigma$ is expressed via $\delta T$ and $a$..

The heat flux through the cylinder base and side surfaces is estimated as $\chi [(\delta T/h_0)a^2+(\delta T/a)h_0a]$. Equating it to the fluctuation of power $V^2a^2\delta\sigma /h_0$ inside the cylinder yields the temperature fluctuation
\begin{equation}\delta T=\frac{V^2\delta\sigma}{\chi}\frac{a^2}{a^2+h_0^2}.\label{eq:dTtoy}\end{equation}
Expressing $\delta\sigma$ from Eq. (\ref{eq:dTtoy}) and substituting it into Eq. (\ref{eq:Stoy}) yields
\begin{equation}\label{eq:Stoy1}
S=\overline{n}a^2\left(\frac{\delta T\chi}{V^2\overline{\sigma}}\right)^2\left(1+\frac{h_0^2}{a^2}\right)^2.\end{equation}

Following the OFM approach, we optimize the exponent $S$ with respect to the fluctuation radius $a$, i. e. $dS/da=0$, which gives $a=h_0$. Substituting $a=h_0$ back into Eq. (\ref{eq:Stoy1}) yields the optimum exponent of probability,
\begin{equation}\label{eq:Stoy2}
S_{opt}= \left(\frac{\delta T}{\delta T_0}\right)^2 \quad {\rm where}\quad\delta T_0\equiv \frac{V^2\overline{\sigma}}{\chi h_0\sqrt{\overline{n}}}
\end{equation}
again to the accuracy of numerical multipliers.

The preexponential is roughly estimated by dividing the entire area into elemental domains of area $h_0^2$ each and noticing that $\exp(-S_{opt})$ describes the probability of a desired fluctuation with temperature excess $\delta T$ in a given domain. Therefore, the concentration of such fluctuations is estimated as $h_0^{-2}\exp(-S_{opt})$.

Two features should be noted. First, OFM concentrates on the exponent of probability, largely neglecting the pre-exponential factors (although they can be estimated as well). Secondly, it optimizes that exponent in order to find the most likely disorder configuration providing the desired fluctuation characteristic of interest. Its applicability is limited to the region of non-overlapping fluctuations.

A possible application of this toy model can might be a system of multiple shunting metal chains formed in dielectric or solid electrolyte films considered for nonvolatile memory; see e.g. Refs. \onlinecite{resistswitch} and references therein.

\subsection{Homogeneous films}\label{sec:homo}

Consider, in the linear approximation, a relatively small temperature fluctuation $\delta T$ in a cylinder region of radius $a$, setting
\begin{equation}
\frac{1}{T}\approx \frac{1}{T_0}-\frac{\delta T}{T_0^2}.
\label{eq:linear}\end{equation}
Neglecting (for simplicity) heat transfer through the cylinder bases and using
$$\delta \sigma = \overline{\sigma}\exp\left(\frac{\delta T\overline{E}}{kT^2}\right),$$
Eq. (\ref{eq:dTtoy}) reduces to the form
\begin{equation}\delta T=\frac{V^2\overline{\sigma}}{\chi}\frac{a^2}{h_0^2}\exp\left(\frac{\delta T\overline{E}}{kT^2}\right).\label{eq:dTtoy1}\end{equation}

For a system in equilibrium, the probability of temperature fluctuation $\delta T$ in volume $\delta V=\pi a^2h_0$ is given by the expression \cite{landau1980} $\exp[-C^{(v)}\delta V(\delta T)^2/kT^2]$ where $C^{(v)}$ is the specific heat per volume. Expressing  $a^2$ from Eq. (\ref{eq:dTtoy1}) gives the equilibrium distribution function $\overline{f}(\delta T)\propto\exp[-S(\delta T)]$ with
\begin{equation}
S(\delta T)=-\frac{\pi C^{(v)}h_0^3\chi}{2kT_0^2V^2\overline{\sigma}}\delta T^3\exp\left(-\frac{\delta T\overline{E}}{kT_0^2}\right).\label{eq:disttoy}\end{equation}

It follows from Eq. (\ref{eq:disttoy}) that the equilibrium distribution is a minimum at $\delta T_c=kT^2/3\overline{E}$ where the product $\delta T^3\exp(-\delta T\overline{E}/kT^2)$ is a maximum. This can be interpreted as a barrier in the system free energy at $\delta T=\delta T_c$: the probability of fluctuations first exponentially decreases as $\delta T$ grows below $\delta T_0$ and then decreases when $\delta T$ exceeds $\delta T_c$. Such a behavior is obviously similar to that known in nucleation phenomena \cite{landau1980,landau2008,kaschiev2000} (where the barrier is a function of the nuclear radius) and small polaron collapse \cite{toyozawa} (where the barrier is a function of dilation). The instability point corresponds to a relatively very small temperature increase $\delta T_c=(kT_0/3\overline{E})T_0\ll T_0$ in systems with high enough activation energies, say, $\delta T_c\lesssim 0.01T_0\sim 3$ centigrade.

Based on that analogy, the exponent of probability of the thermal breakdown is given by $S(\delta T_c)$, that is, to the accuracy of numerical multipliers,
\begin{equation}\label{eq:probabtoy}
S(\delta T_c)=\frac{k^2T_0^3C^{(v)}h_0^3\chi}{\overline{E}^3V^2\overline{\sigma}}.
\end{equation}

Note that the probability exponent optimization here results not in a minimum, but rather a maximum; it may turn into a saddle point in a parameter space of higher dimensionality as will be explicitly shown next. Another conclusion is that a positive feedback alone makes the instability possible regardless of the degree of disorder in the system.
\begin{figure}[t!]
\includegraphics[width=0.45\textwidth]{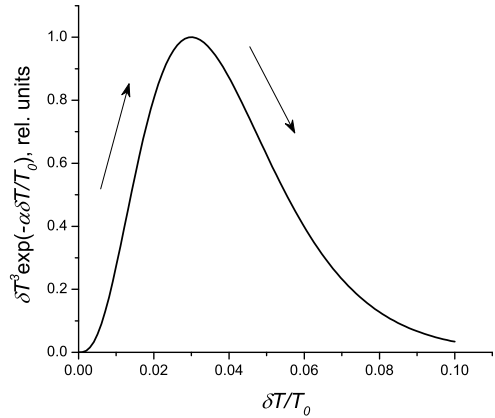}
\caption{Effective barrier for nucleation of hot spots corresponding to the numerical value $\alpha =\overline{E}{kT_0}=10$. Arrows show a pathway of hot spot nucleation. \label{Fig:barrier}}
\end{figure}

\section{Optimum fluctuation method}\label{sec:OFM}
The subtlety of the optimum fluctuation method is in how it treats the disorder induced distribution of temperature $T({\bf r})$ (or wave function for the standard case of energy spectra in systems with random potential energy). Namely, $T({\bf r})$ is considered a smooth 'optimum' function approximating the temperature distribution for the most likely disorder configuration responsible for any desired temperature fluctuation. It remains arbitrary (yet fixed) in the course of the analysis and is determined later by the condition of the maximum of the probability. Such optimization benefits from the known property of variational techniques that any inaccuracy in the trial function translates into a higher order inaccuracy in the corresponding functional.

In what follows we take into account only exponentially strong activation factor ignoring all possible pre-exponentials found with temperature dependent conductivity in semiconductors. This simplification simultaneously determines the accuracy of our analysis where all the pre-exponential factors are replaced with their averages. In particular, this analysis is limited to the case of strong enough fluctuations beyond the linear approximation for $P[\delta T({\bf r})]$.
\subsection{OFM equations}\label{sec:OFMe}
The heat transport equation (\ref{eq:heattr}) can be treated as an extremum of the functional
\begin{equation}
F=\int d^3r\left[\frac{\xi}{2}(\nabla T)^2-P({\bf r})\right]
\label{eq:func1}\end{equation}
where the pre-exponential factor $(E+\overline{E})/T^2$ [generated by variation of $P$ in Eq. (\ref{eq:locpow})] is approximated by its average,
\begin{equation}\label{eq:Theta}
\xi \equiv \chi \langle (\overline{E}+E)/kT^2\rangle.\end{equation} The latter functional can be presented as
\begin{equation}
F=\int d^3r\left[\frac{\xi}{2}(\nabla T)^2-\langle P\rangle\right]-Z
\label{eq:funcheat}\end{equation}
where $T$ depends on coordinates and random variable $Z$ is defined by
\begin{equation}
Z=\int d^3rP^{(1)}({\bf r})
\label{eq:randvarz}\end{equation}

OFM suggests that the dispersion of random variable $Z$ can be found as
\begin{equation}
D =\langle Z^2\rangle =\int\int d^3rd^3r'\langle P^{(1)}({\bf r})P^{(1)}({\bf r'})\rangle
\label{eq:disp}\end{equation}
where the average in the integrand is evaluated under the condition of a fixed (yet arbitrary) function $T({\bf r})$. The integral in Eq. (\ref{eq:randvarz}) contains a large number of random contributions. Therefore, according to the central limit theorem, $Z$ is described by Gaussian statistics, i. e. its probabilistic distribution
\begin{equation}
g(Z)\propto \exp[-S(Z)],\quad S=\frac{Z^2}{2D}.
\label{eq:Gauss}\end{equation}

The maximum probability fluctuation corresponds a stationary point of $S(Z)$ under the additional condition of Eq. (\ref{eq:funcheat}). Finding such a conditional extremum is tantamount to finding an unconditional extremum of a functional
\begin{equation}\Phi =\frac{Z^2}{2D}-\lambda F
\label{eq:uncond}\end{equation}
where $\lambda$ is the undetermined Lagrange multiplier. $\lambda$ is then found from the additional condition of a certain predetermined  maximum temperature in the the optimum fluctuation region.

The functional $\Phi$ must be optimized with respect to the disorder configuration $E({\bf r})$ and the field $T({\bf r})$. Because the former appears only with the integral $Z$, the optimization can be more conveniently conducted with respect to $Z$ and $T({\bf r})$. The corresponding equations are
\begin{equation}
\frac{Z}{D}+\lambda =0
\label{eq:optz}\end{equation}
and
\begin{eqnarray}
&&-\frac{Z^2}{2D^2}\frac{\delta D}{\delta T}+\lambda\xi \nabla ^2T+ \nonumber \\ &&\lambda P_0\left(\frac{{\overline E}}{kT^2}-\frac{\langle E^2\rangle}{k^2T^3}\right)\exp\left(\frac{{\langle
E^2\rangle}}{2k^2T^2}\right)=0.
\label{eq:OFM}\end{eqnarray}
Here we have taken into account a known property \cite{landau1980}
$$\langle \exp(E/kT)\rangle =\exp[\langle (E/kT)^2\rangle /2]$$
for a Gaussian random variable $E/kT$. Using Gaussian statistics in combination with the concept of thermally activated current assumes the inequality
\begin{equation}\label{eq:weakfluc}
\frac{{\overline E}}{kT}\gg\frac{\langle E^2\rangle}{k^2T^2}.
\end{equation}
Allowing the opposite inequality would lead to the physically unacceptable feature that the typical fluctuation current exponentially decreases with temperature.

Substituting Eq. (\ref{eq:optz}) into Eqs. (\ref{eq:Gauss}) and (\ref{eq:OFM}) yields the equations determining the optimum fluctuation temperature field $T({\bf r})$ and its corresponding probability exponent,
\begin{eqnarray}\label{eq:OFM1}
&&-\frac{\lambda D}{2}\frac{\delta D}{\delta T}+\xi \nabla ^2T+ \nonumber \\&& P_0\left(\frac{{\overline E}}{kT^2}-\frac{\langle E^2\rangle}{k^2T^3}\right)\exp\left(\frac{{\langle
E^2\rangle}}{2k^2T^2}\right)=0,
\end{eqnarray}
\begin{equation}
S=\frac{D\lambda ^2}{2}.
\label{eq:SvsD}\end{equation}

To evaluate $\delta D/\delta T$ that is the variational derivative of the integrand in Eq. (\ref{eq:disp}) we use again the property of averaging of a Gaussian random variable $E({\bf r})$. The integrand in Eq. (\ref{eq:disp}) becomes
\begin{equation}
P_0^2\exp\left[\frac{\langle E^2\rangle}{(kT)^2}\right]\int d^3r'\left\{\exp\left[\frac{\langle E({\bf r})E({\bf r'})\rangle}{k^2T({\bf r})T({\bf r'})}\right]-1\right\}.\nonumber
\label{eq:PPcor}\end{equation}
For the case of delta correlated disorder in Eq. (\ref{eq:corfun}), the latter expression can be approximated as
\begin{equation}
P_0^2sh_0\exp\left[\frac{2B}{(kT)^2}\right].
\label{eq:PPcor1}\end{equation}

Substituting the result of differentiation [together with Eq. (\ref{eq:optz})] into Eq. (\ref{eq:OFM}) leads to a closed form single equation for the optimum fluctuation $T({\bf r})$. That equation is not very useful practically because of its rather complex form . The problem becomes easier when presented in the form of functional subject to direct optimization with respect to $T({\bf r})$. That functional is given by
\begin{equation}J=\int d^3rF[T({\bf r})]
\label{eq:funcJ}\end{equation}
where
\begin{eqnarray}
F=\frac{\xi}{2}(\nabla T)^2-P_0\exp\left[\frac{B}{2(kT)^2}\right]-
\lambda P_0^2v\exp\left[\frac{2B}{(kT)^2}\right].
\label{eq:funcinJ}
\end{eqnarray}
Note that, to the accuracy of the factor of $-\lambda$, the third term in the functional $J$ [corresponding to the third term in Eq. (\ref{eq:funcinJ})] is twice the probability exponent $S$.

\subsection{OFM saddle points}\label{sec:OFMt}
While optimization of functional $J$ remains to be implemented, the nature of its stationary points can be determined already here. Assuming a trial function $T=T(r/a)$ and changing variable $r\rightarrow r/a$, $J$ can be presented in the form
$$J=J_1+a^2J_2$$
where $J_1$ and $J_2$ do not depend on $a$. Treating $a^2$ as a variational parameter, leads to the conclusion that $d^2J/d(a^2)^2=0$ at the stationary points where $J_2=0$. Hence, they represent inflection points rather than minima. In a higher dimension parameter space including the temperature fluctuation amplitude, these points can only be saddles.

The saddle point solutions require a different interpretation of OFM results. From the physical standpoint, some (but not all) of their related configurations should appear with certainty, i. e. with $S=0$, since they are not steady state, and thus are to be passed inevitably sooner or later. From that perspective, they are similar to the barriers of classical nucleation theory \cite{landau1980,landau2008,kaschiev2000} or small radius acoustic polaron formation. \cite{toyozawa} For example, the OFM saddle points in the surface $J(a,T)$ can physically describe critical radii $a(T)$ separating the regions of spontaneous decay from that of spontaneous growth of fluctuations. This similarity to the nucleation theory will be made explicit in Sec. \ref{sec:variational}.

Note that the fact of probability exponent $S$ vanishing at the OFM saddle points, does not compromise OFM as long as the corresponding fluctuations remain strongly localized and do not overlap. The latter conditions do not necessarily invoke $S\gg 1$ (unlike the conclusion of Sec. \ref{sec:toymodel} where all the fluctuations simultaneously coexist), since the saddle point events are not steady state taking place at different time instances.

Consider the configurational probability exponent $S$ in a certain proximity of a saddle point $S=0$. We denote $\delta T_0({\bf r})$ the temperature distribution in the optimum fluctuation corresponding to $S=0$. If the optimum fluctuation $\delta T({\bf r})$ is different from $\delta T_0({\bf r})$, one can extend
\begin{equation}
S=\int d^3r\left(\frac{\delta ^2S}{2\delta T^2}\right)_0[\delta T({\bf r}) -\delta T_{\beta}({\bf r})]^2,
\label{eq:SdT}\end{equation}
where the integrand is positive. The equilibrium distribution function of such fluctuations is given by
\begin{equation}\overline{f}(\delta T)=\overline{f}_{0}\exp\left\{-\frac{C^{(v)}}{2kT_0^2}\int d^3r[\delta T ({\bf r})]^2-S\right\}\label{eq:probtime}\end{equation}
Here $\overline{f}_{0}$ is the preexponential factor and we have taken into account the expression for the probability of equilibrium temperature fluctuation $\delta T$ in volume $\delta V$ mentioned in Sec. \ref{sec:homo}.

It is seen from Eq. (\ref{eq:probtime}) that $\overline{f}$ is a minimum at some $\delta T$ different from $\delta T_0$. Following the Fokker-Planck approach to nucleation (Zeldovich' theory; see e. g. Chapter XII in Ref. \onlinecite{landau2008}) and in agreement with the qualitative analysis in Sec. \ref{sec:homo}, that minimum determines the nucleation barrier and rate. This approach will be implemented in Sec. \ref{sec:FPE} below upon determining the parameters of OFM solutions $\delta T({\bf r})$.

\section{Direct variational procedure}\label{sec:variational}
\subsection{Trial function and functional}\label{sec:trial}
Here we implement a direct variational procedure of optimization of the functional $J$ using the simplest trial function
\begin{equation}
\frac{\delta T}{T_0} =\theta \left(1-\frac{r}{\tilde{a}h} \right)\left(1-\frac{z}{h} \right)\quad {\rm when} \quad \delta T>0
\label{eq:trial}\end{equation}
that is zero outside of the domain $r<\tilde{a}h$, $z<h$. Here $r$ and $z$ are the radial and transversal (across the film) coordinates. $\theta $ and $\tilde{a}$ are the two variational parameters, defined as being dimensionless to make the resulting equations more compact. In particular, $\theta$ is the amplitude excess temperature in fluctuation measured in the units of the average temperature $T_0$, and $\tilde{a}$ has the meaning of the fluctuation radius measured in the units of structure thickness $h$.

Note that integration over the transversal ($z$) coordinate extends over the entire structure thickness ($h$) for the first term in Eq. (\ref{eq:funcinJ}), while the second and third terms must be integrated only over the active layer thickness ($h_0\ll h$) where the power is generated. Also, we note that the constraint $t=0$ at $z=h$ correctly reflects the boundary condition of a constant temperature at the interface (see Fig. \ref{Fig:sample}). Furthermore, we assume fluctuation to be relatively small, allowing the linearization in Eq. (\ref{eq:linear}).

Substituting Eq. (\ref{eq:trial}) and carrying out the integration reduces $J$ to the form
\begin{equation}\label{eq:Jspec}
\frac{12\alpha ^2J}{\xi T_0^2\pi h}= \left( \tilde{a}^2 +2 \right) x^2
- \beta \tilde{a}^2\Phi (x)-\lambda \beta ^{'}\beta \tilde{a}^2\Phi\left(x\frac{\alpha^{'}}{\alpha} \right)
\end{equation}
where
\begin{equation}\label{eq:Ft}
x=\alpha\theta \quad {\rm and}\quad \Phi(x)=\frac{\exp(x)- x -1}{x^2}.
\end{equation}
Here we have introduced the parameters defined as
$$\alpha =\frac{\overline{E}}{kT_0}-\frac{B}{(kT_0)^2},\quad \alpha ^{'}=2\left[\alpha -\frac{B}{(kT_0)^2}\right]\approx 2\alpha ,$$
$$\beta =\frac{24hh_0P_0\alpha^2}{\xi T_0^2}\exp\left[\frac{B}{2(kT_0)^2}\right], \beta ^{'}=P_0v\exp\left[\frac{3B}{2(kT_0)^2}\right].$$
The inequality in Eq. (\ref{eq:weakfluc}) limits them to $\alpha \gg 1$.
In integrating over $z$ in Eq. (\ref{eq:Jspec}), we have assumed a practically important case when the semiconductor layer is very thin, $\alpha \theta h_0/h \ll 1$, and calculations are simpler.

Because eventually we consider $\theta =x/\alpha$ an independent {\it given} variable, the optimization conditions $\partial J/\partial \tilde{a}^2=0$ and $\partial J/\partial x=0$ must be used to solve for $\tilde{a}^2$ and $\lambda$. In agreement with the conclusion of Sec. \ref{sec:OFM}, the stationary points found from the optimization are saddle points. This is seen from the sign of the determinant
$$\frac{\partial^2 J}{(\partial \tilde{a}^2)^2}\frac{\partial^2 J}{(\partial \theta)^2}-\left[\frac{\partial^2 J}{(\partial \tilde{a}^2) \partial \theta}\right]^2<0$$
identifying the stationary points as saddles. \cite{korn2000}
\subsection{Regional approximations}\label{sec:regional}
Consider the results of optimization of the functional $J$ for three complimentary regions.
\subsubsection{Weak fluctuations, $x\ll 1$}\label{sec:weak}
Assuming $x\ll 1$ reduces $\Phi (x)$ in Eq. (\ref{eq:Jspec}) to $\Phi (x)\approx 1/2+x/6+x^2/24$, which significantly simplifies the optimization. This leads to the physically unacceptable solution with $\tilde{a}^2=-32/(8+\beta )<0$.
\subsubsection{Moderate fluctuations, $x\sim 1$}\label{sec:mod}
It is straightforward to verify that the interpolation $\Phi (x)=1/2+x^2/6$ holds to the accuracy of several percent for intermediate $x\leq 4$. Using that interpolation, the optimization of $J$ results in the physically inconsistent solution as well, $\tilde{a}^2=-[12+16(\alpha\theta )^2]/(6+3\beta )$.

\begin{figure}[tb]
\includegraphics[width=0.46\textwidth]{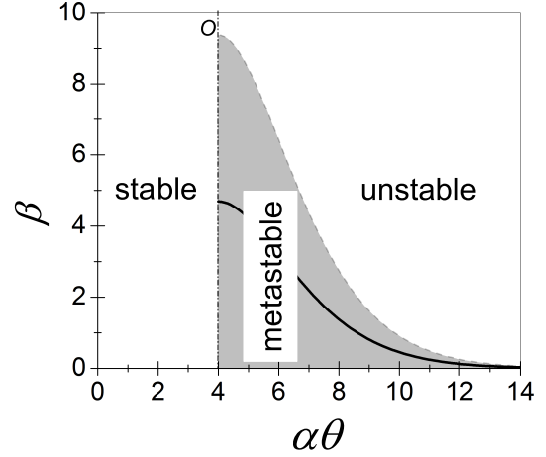}
\caption{Phase diagram for a thin film structure with transversal current vs. power density (parameter $\beta$) and local temperature increase (parameter $\alpha\theta$).  Region to the left of the line $\alpha\theta =4$ represents the stable phase where local temperature fluctuations decay making thermal breakdown impossible. The gray colored region below the line of solution of Eq. (\ref{eq:transc}), represents metastable state corresponding to the saddle points, through which thermal breakdown nucleates locally. The solid curve in that region is a solution of Eq. (\ref{eq:ridge}); it corresponds to the most likely nucleation events, for which $S=0$ in Eq. (\ref{eq:results}). The region above the line of solution of Eq. (\ref{eq:transc}) represents the globally unstable state of the system.  \label{Fig:ranges}}
\end{figure}
\subsubsection{Strong fluctuations, $x\gg 1$}
Acceptable solutions with $a^2>0$ exist in the case of $\alpha\theta\gg 1$ (and yet $\alpha \theta h_0/h \ll 1$) where one can approximate $\Phi (x)=\exp(x^2)/x^2$. This yields
\begin{eqnarray}\label{eq:results}
\lambda &=&\frac{(\alpha\theta)^4-\beta\exp(\alpha\theta)}{\beta\beta ^{'}\exp(2\alpha\theta)},\nonumber\\
\tilde{a}^2&=&\frac{4(\alpha\theta)^3}{2(\alpha\theta)^4-\beta\exp (\alpha\theta)},\label{eq:a^2}\\
S&=&S_0\frac{\theta [(\alpha\theta)^4-\beta\exp(\alpha\theta)]^2\exp(-2\alpha\theta)}{2(\alpha\theta)^4-\beta\exp(\alpha\theta)}\nonumber\\
\end{eqnarray}
where
\begin{equation}\label{eq:S0}
S_0\equiv \frac{\pi (\xi T_0^2)^2\exp[-2B/(kT_0)^2]}{288P_0^2vh_0}\end{equation}
and
\begin{align}\label{eq:rootsCond}
\theta _{c1}<\theta <\theta _{c2},
\end{align}
with $t_{c1}$ and $t_{c2}$ being the two solutions of the transcendental equation
\begin{equation}2(\alpha\theta )^4-\beta\exp(\alpha\theta )=0.\label{eq:transc}\end{equation}

The condition
\begin{equation}(\alpha\theta)^4-\beta\exp(\alpha\theta)=0\label{eq:ridge}\end{equation}
describes the points where $S=0$ and thus nucleation of hot spots takes place, according to the discussion in Sec. \ref{sec:OFMt}. These points all fall within the domain of physically acceptable solutions in Eq. (\ref{eq:rootsCond}). Also, it follows from comparison of Eqs. (\ref{eq:transc}) and (\ref{eq:ridge}) that the radii of the corresponding stationary fluctuation states remain finite as required by OFM.

Because $(\alpha\theta )^4\exp(-\alpha\theta )$ is a maximum at $\alpha\theta =4$, Eq. (\ref{eq:ridge}) has solutions when
\begin{equation}\beta \leq \beta _c=\left(\frac{4}{[e]}\right)^4\approx 4.7\label{eq:betac}\end{equation}
where $[e]$ stands for the base of natural logarithms. Close to that threshold value, the dependence $t(\beta )$ takes the form
\begin{equation}\alpha\theta\approx\alpha\theta _0 =4+\sqrt{\beta _c-\beta}\quad {\rm when}\quad \beta _c-\beta\ll 1.\label{eq:proxbeta}\end{equation}
Another branch of $\alpha\theta$ with the minus sign before the square root is ignored as belonging to the moderate fluctuation regime.

Alternatively, one gets from Eq. (\ref{eq:ridge}),
\begin{equation}\alpha\theta\approx \alpha \theta _0=\ln(1/\beta )\gg 1\quad {\rm when}\quad \beta\ll \beta _c.\label{eq:lowpow}\end{equation} This behavior corresponding to the far right part of the solid curve in Fig. \ref{Fig:ranges} describes the low power regime.

\subsection{Phase diagram}\label{sec:phased}
The complementary region to the left of the line $\alpha\theta =4$ in Fig. \ref{Fig:ranges} was characterized by the physically unacceptable solutions with $\tilde{a}^2<0$ (see Sec. \ref{sec:weak} and \ref{sec:mod}). Here, we argue that that region represents the state where the system remains stable with respect to thermal fluctuations. A proof is achieved by including in the above analysis the term $-C\delta T/\tau$ from Eq. (\ref{eq:transheattr}) describing the temporal behavior of fluctuation. It is straightforward to see that the unacceptable negative $\tilde{a}^2$ turn positive when $\tau >0$, i. e. the corresponding fluctuations decay.

Alternatively, for the region above the curve $\beta =2(\alpha\theta)^4\exp(-x)$, adding the term with negative relaxation time $\tau <0$ allows for positive $\tilde{a}^2$. Therefore, the states in that region are globally unstable, i. e. they evolve into highly conductive high temperature states without any barrier. This is qualitatively similar to the phase transition scenario of spinodal decomposition, \cite{cahn} which is not described in the OFM framework.

Note the triple point $O$ at ($\beta =\beta _c$, $\alpha\theta =4$) in Fig. \ref{Fig:ranges} where all three phases coexist. It is straightforward to show that fluctuations $\delta \theta$ become increasingly strong in its proximity where
\begin{equation}\label{eq:Sexpand} S=-\frac{S_0\beta ^2\alpha}{(\alpha\theta_0)^5}(\alpha\theta_0 -4)^2(\delta\theta )^2\end{equation}
and $|\delta\theta |=|\theta -\theta _0|\ll \theta _0$. That property is similar as well to that of the standard phase transition phase equilibria.\cite{landau1980}

\subsection{Approximation of classical nucleation theory}\label{sec:CNT}
The approximation of classical nucleation theory implies a narrow boundary region between the two phases and its related concept of surface energy. It can be attempted in the current framework by choosing a trial function
\begin{equation}
\frac{\delta T}{T_0} =\theta\left\{\begin{array}{lll}1\quad \textrm{when}\quad r<a,\\
(a+d-r)/d\quad \textrm{when} \quad a<r<a+d,\\
0\quad \textrm{when} \quad r>a+d\end{array}\right.
\label{eq:trial1}\end{equation}
with $d\ll a$. As a result, the gradient term in Eq. (\ref{eq:funcinJ}) is determined by the contribution from a narrow layer of width $d$ analogous to nucleus interfacial energy in functional $J$ of Eq. (\ref{eq:funcJ}). The procedure of optimization becomes even simpler than that based on the trial function of Eq. (\ref{eq:trial}). Omitting the details, the result is that the functional $J$ has no stationary points when $d\ll a$. Hence, the approximation of interfacial energy does not apply to the case under consideration; the function in Eq. (\ref{eq:trial}) remans more adequate.

\section{Steady state transition rate}\label{sec:FPE}
Consider the probability of thermal breakdown at a given power density $P_0$ described in terms of the parameter $\beta <\beta _c$.
Using $\delta T(r,z)$ from Eq. (\ref{eq:trial}) and expressions for $\tilde{a}$ and $S$ from Eq. (\ref{eq:results}), the equilibrium distribution function becomes
\begin{equation}\overline{f}(\theta)=f_{0}\exp\left[-\frac{\pi C^{(v)}h^2h_0\theta ^2\tilde{a}^2(\alpha\theta )}{3k}-S(\alpha\theta )\right].
\label{eq:probtime1}\end{equation}

$S(\alpha\theta )$ is a maximum, $S=0$, at the line shown in Fig. \ref{Fig:ranges} and increases towards the boundary $\alpha\theta =4$. However, given realistic parameters (see Sec. \ref{sec:disc}) that increase is not nearly as significant as the increase of the first term in the exponent in Eq. (\ref{eq:probtime1}). As a result, $\overline{f}(\theta)$ has a sharp minimum at $\alpha\theta \approx 4$.

Following the known approach of nucleation theory \cite{landau2008} (mentioned in Sec. \ref{sec:OFMt} above) consider a stationary Fokker-Planck equation
\begin{equation}
j=-B\frac{\partial f}{\partial \theta}+Af=const
\label{eq:FPE}\end{equation}
for the 'kinetic' temperature distribution function $f(\theta )$. Here $j$ is the flux in the temperature fluctuation ($\theta$) space, $D$ is the diffusion coefficient in that space; $A$ is connected with $D$ by a relationship which follows from the fact that $j=0$ for the equilibrium distribution $f=\overline{f}$. Using the latter enables one to present the flux as
$j=-B\overline{f}(\partial /\partial \theta)(f/\overline{f})$, and, hence,
$f/\overline{f}=-s\int d\theta /B\overline{f}+const.$
Finally, applying the boundary conditions $f\rightarrow 0$ when $t\rightarrow\infty$ and $f= \overline{f}$ when $\theta =0$, yields
\begin{equation}
\frac{1}{j}=\int_0^{\infty}\frac{d\theta }{B\overline{f}}.\label{eq:sfinal}\end{equation}
The integral is determined by a narrow proximity of the minimum of $\overline{f}$ that gives the exponent of the transition rate.

To roughly evaluate the preexponential factor (without any knowledge of $D$) one can divide the entire area into a set of cells of characteristic linear size of the optimum fluctuation $\tilde{a}h$. Then the preexponential must be of the order of the rate of temperature variations $\kappa /(\tilde{a}h)^2$ in a cell where $\kappa$ is the thermal diffusivity. This yields the steady state nucleation rate (cm$^{-2}$s$^{-1}$),
\begin{equation}\label{eq:nucratefinal}
j\sim\frac{16\kappa}{h^4}\exp\left[-\frac{\pi C^{(v)}h^2h_0\theta ^2\tilde{a}^2(4)}{3k}-S(
4)\right]
\end{equation}
where $\tilde{a}^2(4)\equiv \tilde{a}^2(\alpha\theta =4)$ and  $S(4)\equiv S(\alpha\theta =4)$ are given in Eq. (\ref{eq:results}) The power density enters this result through the parameter $\beta$ in Eq. (\ref{eq:S0}).

This result becomes more explicit for the case of low enough power when $\beta\exp(4)\ll 4$ in Eq. (\ref{eq:results}) and the absolute value of the exponent in Eq. (\ref{eq:nucratefinal}) is estimated as
\begin{equation}
S\approx 8\frac{C^{(v)}h^2h_0}{\alpha ^2k}+7\cdot 10^{-6}\frac{(\xi T_0^2)^2\exp[-2B/(kT_0)^2]}{\alpha h_0^2sP_0^2}.
\label{eq:expnucl}\end{equation}
This is similar to the exponent in Eq. (\ref{eq:probabtoy}) emphasizing the important role of specific heat and rapidly decreasing with the power density. However it has a distinct feature of a lower boundary beyond which it cannot be further reduced even for very high power densities. It should be remembered however that high enough power densities are conducive to a different type of instability similar to the spinodal decomposition transformations as reflected in Fig. \ref{Fig:ranges}.

\section{Discussion and conclusions}\label{sec:disc}
\subsection{Numerical estimates}\label{sec:charpar}
Assuming the typical semiconductor values, \cite{sze} one gets
$\chi\sim 1$ W/cm-grad and $\overline{E}/T\sim 10-100$ for activation energies $\overline{E}\sim 1$ eV and $T\sim 100-500$ $^o$K.
This yields $\xi\sim 1-100$ W/cm-grad$^2$, $\alpha\sim 10-100$.

For geometrical parameters, it is natural to assume $h_0\sim 1$ $\mu$m, $s\sim 1$ $\mu$m$^2$, and $h\sim 10^{-4}-10^{-1}$ cm.
The current density in the range from 1 $\mu$A/cm$^2$ to 1 A/cm$^2$ and electric fields ${\cal E}\sim 10^3-10^5$ V/cm are used in many device operations. The corresponding power densities are in the range from 1 mW/cm$^3$  to $10^5$ W/cm$^3$. The fluctuation strengths exponent $\exp[-2B/(kT_0)^2]$ can be evaluated as $\sim 0.001 - 1$ based on the observations of transversal currents through nonuniform Schottky barriers and thin film photovoltaics. \cite{schottky} Finally, we use the thermodynamic parameters $C\sim 0.1 - 1$ J/sm$^3$-$^o$C and $\kappa\sim 0.1-1$ cm$^2$/s.
With the above parameters, the preexponential factor in Eq. (\ref{eq:nucratefinal}) is estimated as $\sim 10^{5}-10^{13}$ cm$^{-2}$s$^{-1}$. Given that preexponential, the exponent in Eqs.  (\ref{eq:nucratefinal}) and (\ref{eq:expnucl}) can be then within the range of experimentally important nucleation rates only for micron or sub-micron thin devices. Assuming greater thickness, say, $h\gtrsim 1$ mm makes the thermodynamic term proportional to $C$ large enough to practically rule out the possibility of thermal breakdown mechanism under consideration.

However, semiconductor devices of modern electronics are often 10-100 nm thick (unless intended thermal sinks are used), and for them the thermodynamic fluctuation term in the exponent is not too large. For such structures, the second term in the nucleation rate exponents can be not terminally large for powers in the range $P_0 \gtrsim 100$ W/cm$^3$. Overall, this makes the above considered mechanism realistic for structures in submicron region.

Finally, the minimum power density corresponding to the critical value of $\beta$ in Eq. (\ref{eq:betac}), above which the nucleation mechanism turns into that of global instability, can be estimated as $P_0\gtrsim 10^{11}$ W/cm$^3$. This range of power density is above practically all types of modern semiconductor devices, except maybe some cases of power electronics.

\subsection{Discussion}\label{sec:gendis}

\begin{figure}[tb]
\includegraphics[width=0.42\textwidth]{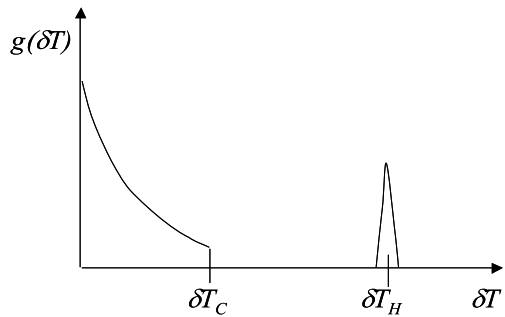}
\caption{Probability $g(\delta T)$ of hot spots vs. their excess temperature $\delta T$. The Gaussian tail at low $\delta T$ is described in Sec. \ref{sec:homo}. The critical overheat $\delta T_c$ corresponds to the condition $\alpha\Theta =4$ illustrated in Fig. \ref{Fig:ranges}. The high temperature peak at $\delta T_H$ is determined by the processes of saturation of activated conduction and inter-spot interactions as explained in Sec. \ref{sec:gendis}; its width is due to disorder effects.  \label{Fig:prob}}
\end{figure}
The above consideration is limited to a basic instability triggered by Joule heat in combination with activated conduction. The instability is predicted to start under insignificant local overheats of several degrees. However, this analysis does not address the final parameters to which the instability can grow.

The 'stabilized' temperature excess $\delta T_H$ in the developed filament (beyond the present theory framework) can be rather substantial. As pointed in Ref. \onlinecite{subashiev1987}, it can belong in the temperature range where the activated conduction saturates. That high temperature limit should not be mixed with the above predicted transition temperature excess, $\delta T_c\approx 4kT^2/\overline{E}\ll \delta T_H$ (corresponding to $\alpha\Theta\approx 4$), starting from which the instability evolves. This is illustrated in Fig. \ref{Fig:prob}.

Furthermore, it is conceivable that the steady state high temperature local overheat $\delta T_H$ cannot be determined by any extension of the present theory limited to noninteracting hot spots, even if activated conduction is allowed to saturate. The concentration of steady state hot spots at $\delta  T_H$ can significantly depend on their interaction. Indeed, the present theory predicts (Sec. \ref{sec:FPE}) that even at arbitrarily however low rates, the above described instabilities will keep developing (maybe beyond the practically significant time intervals) to take over the entire structure area. This contradictory prediction is not unique of the system under consideration. It is known in the theory of phase transition where the nucleation stage is limited by various inter-nucleus interactions, such as competition for material, elastic stresses, etc. Similarly limiting interactions here will include competition of hot spots for the electric current, thermal fields by other filaments, etc. This analogy leads to the prediction of the growth and ripening stages of thermal breakdown kinetics, similar to that of the standard phase transitions; \cite{landau2008} a theory of such later stages of hot spot transformation remains to be developed.

While not related to structural transformations, the predicted local temperature increase can accelerate such transformations leading to permanent failures in the form of conducting  pathways. Therefore, this mechanism can serve as a precursor to permanent structural failures.
From that point of view, the above results on low temperature thermal breakdowns $\delta T_c\ll T$ point at high sensitivity of the fatal failure probability to the activation energy of conductivity and thermodynamic variables, particularly, specific heat, thickness, and thermal insulation.

The role of inactive (thermally insulating) layers exponentially reducing the thermal breakdown rates is due to the filament diameter increase with its length.  As a result the thermal gradient in radial direction decreases suppressing the instability rate. This is consistent with the known practical solutions using substantial heat sinks attached to with submicron electronic devices in order to minimize their failure rates.

A more theoretical comment is in order regarding the relevance of the above OFM modification aimed at `nontraditional' saddle type of stationary points. The underlying motivation was to relate localized temperature fluctuations with other  known localized states in disordered systems. However the same basic equations as derived in Sec. \ref{sec:OFM} could be obtained in the framework of instanton approach suitable for theoretical description of nucleation. \cite{dykman1,dykman2,langer} That approach would start with the time dependent heat transfer equation leading to the variational problem for the exponent of probability $\exp[-R(T,t)]$ where $t$ is time and $R$ is related to the functional in Eq. (\ref{eq:func1}), $R\propto \int^tF[T(t)]dt$. $F$ remains a random functional to be additionally optimized to maximize the probability. That reduces the conditional variational problem for $R$ to that of unconditional extremum in Eq. (\ref{eq:uncond}) yielding final expressions of OFM in Eq. (\ref{eq:OFM1}).

The above theory has the following limitations. (1) The assumption of fixed voltage $V$ across the film implying that the current $I$ through the filament must be small enough, $IR_{sh}\ll V$ where $R_{sh}$ is the sheet resistance of the conductive electrodes. (2) Simplification of uniform thermal conductivity may have noticeable quantitative ramifications, yet can hardly change the qualitative predictions. (3) The approximation of $\delta$-correlated disorder, according to which the transversal conductivity must fluctuate across the distances smaller than the filament radius. The opposite regime of strongly correlated disorder can be readily described by the above results reduced to the case of homogeneous structures, in which then consider $P_0$  or $\overline{\sigma}$ as a random quantity varying over distances greater than the filament radius. (4) The optimum fluctuation method per se with accuracy limited to the probability exponent. (5) Inaccuracy of the direct variational procedure with a simplistic trial function remains unknown. Based on many similar examples, one can expect the results to be semi-quantitatively correct. (6) Limitation of small temperature fluctuations $\alpha \theta h_0/h \ll 1$, remains self-consistent as long as it is consistent with the final results for $\theta$ as it takes place in the above.
\subsection{Conclusions}\label{sec:conc}
The following was shown.\\
(i) Thin film semiconductor structures with activated transversal conduction are unstable with respect to reversible thermal breakdowns in the form of hot spots and their related current filaments.\\
(ii) The instabilities evolve in a manner of phase transitions by either nucleation (at not too high power densities) or absolute instability similar to spinodal decomposition (above certain critical power density).\\
(iii) The optimum fluctuation method can be modified to describe the saddle points, through which such transitions occur.\\
(iv) The instabilities start with finite local temperature fluctuations that are smaller than the average temperature $T_0$ by the factor of $kT_0/\overline{E}$ with $\overline{E}$ being the average activation energy of electric conduction. The initial fluctuation radii are by the same factor smaller than the structure thickness.\\
(v) The stable, metastable, and unstable phases of a thermally uniform system form a diagram (in variables power density -- temperature) similar to the standard phase diagrams of phase equilibria, in particular, with fluctuations diverging towards the triple point.\\
(vi) The steady state nucleation rate of hot spots exponentially depends on the material parameters, system geometry, and disorder strength.

The author hopes that this consideration can form a theoretical basis to analyze system failures in various structures of modern thin film devices; specific examples will be presented elsewhere.
\acknowledgements
This work was performed under the auspice of the NSF award No.  1066749.
Discussions with I. V. Karpov, A. V. Subashiev, A. Vasko, and K. Wieland are greatly appreciated.

\end{document}